\documentclass[useAMS,usenatbib]{mn2e}
\usepackage{graphicx}
\usepackage{array}
\usepackage{color}
\usepackage{natbib}
\bibpunct{(}{)}{;}{a}{,}{,}

\newcommand{\degree}{\ensuremath{^\circ}}

\title[Magnetic field geometry of LBN 437]{Magnetic fields in cometary globules - IV. LBN 437}
\author[Soam et.~al]{Soam, A.$^{1}$\thanks{email:{archana@aries.res.in}}, 
Maheswar, G.$^{1}$,
Bhatt,H. C.$^{2}$,
Chang Won Lee$^{3}$,
Ramaprakash, A. N.$^{4}$
\\
$^{1}$ Aryabhatta Research Institute of Observational Sciences (ARIES), Nainital 263002, India\\
$^{2}$ Indian Institute of Astrophysics, Kormangala (IIA), Bangalore 560034, India.\\
$^{3}$ Korea Astronomy $\&$ Space Science Institute (KASI), 776 Daedeokdae-ro, Yuseong-gu, Daejeon 305-348, Republic of Korea.\\
$^{4}$ Inter-University Centre for Astronomy and Astrophysics (IUCAA), Ganeshkhind, Pune 411007, India\\}

\begin{document}
\date{Accepted------}

\pagerange{\pageref{firstpage}--\pageref{lastpage}} \pubyear{--}

\maketitle

\label{firstpage}

\begin{abstract}
We present results of our $R-$band polarimetry of a cometary globule, LBN 437 (Gal96-15, $\ell$ $=$ 96$\degree$, \textit{b} $=-15\degree$), to study magnetic field geometry of the cloud. We estimated a distance of $360\pm65$ pc to LBN 437 (also one additional cloud, CB 238) using near-IR photometric method. Foreground contribution to the observed polarisation values was subtracted by making polarimetric observations of stars that are located in the direction of the cloud and with known distances from the Hipparcos parallax measurements. The magnetic field geometry of LBN 437 is found to follow the curved shape of the globule head. This could be due to the drag that the magnetic field lines could have experienced because of the ionisation radiation from the same exciting source that caused the cometary shape of the cloud. The orientation of the outflow from the Herbig A4e star, LkH$\alpha$ 233 (or V375 Lac), located at the head of LBN 437, is found to be parallel to both the initial (prior to the ionising source was turned on) ambient magnetic field (inferred from a star HD 214243 located just in front of the cloud) and the Galactic plane.
  
\end{abstract}


\section{Introduction}

Cometary globules (CGs) are molecular clouds that show a compact bright-rimmed head and a faint tail geometry that extends from the head and points away from a nearby photoionising source. These objects were first noted by \citet{1976MNRAS.175P..19H} on SERC IIIaJ Sky Survey plates. A group of $\sim30$ CGs, the largest of such system, has been identified in the Gum nebula \citep{1976MNRAS.177P..69S, 1983A&A...117..183R, 1983ApL....23..119Z}. Cometary shaped clouds are often found associated with HII regions and OB stars \citep{1976MNRAS.175P..19H, 1976MNRAS.177P..69S, 1980ApJ...240...84S, 1983ApL....23..119Z, 1983A&A...117..183R, 1986Ap.....24..119G, 1991ApJS...77...59S, 1992ApJ...390L..13B, 1998PASA...15...91O, 2007MNRAS.379.1237M}. A number of CGs, however, are found to be relatively isolated, for example CG 12 \citep{1977MNRAS.180..709W, 2004MNRAS.355.1272M}. CGs are considered to be a subset of Bok globules with a size range of $\sim0.1-1.0$ pc \citep{1976MNRAS.175P..19H, 1983ApL....23..119Z}. They exhibit high densities, $10^{4}$-$10^{5}$ $cm^{-3}$ \citep{1994ApJ...433...96V, 1995MNRAS.276.1067B, 2007A&A...466..191H} with mass range of 10-100 M$_{\odot}$ \citep{1994A&A...289..559L, 2007A&A...466..191H} and with a kinetic temperature of $\sim15-35$ K \citep{1990A&A...233..197H, 1991psfe.conf..287C, 1993A&A...274L..33W, 1994A&A...290..235O, 1995A&A...293..493G, 2007A&A...466..191H}. There is evidence for ongoing low mass star formation in a number of CGs  \citep[e.g., ][]{1977MNRAS.180..709W, 1983A&A...117..183R, 1983MNRAS.203..215B, 1984A&A...139..135P, 1998AJ....116.1376S, 2002AJ....123.2597O, 2004A&A...416..677A, 2005AJ....129.1564K, 2007ApJ...654..316G, 2007MNRAS.379.1237M, 2008ApJ...673..331G, 2008AJ....135.2323I, 2009ApJ...706..896M, 2012AJ....143...61N, 2013AJ....145...15R, 2013A&A...550A..83M}. 

Processes involved in the formation and evolution of CGs have  been discussed by a number of authors \citep[e.g., ][]{1989ApJ...346..735B, 1990ApJ...354..529B, 1992IAUS..150..133S, 1994A&A...289..559L, 1995A&A...301..522L, 2001MNRAS.327..788W, 2003csss...12..799K, 2006MNRAS.369..143M, 2007IAUS..237..450M}. Pre-existing small, dense cores distributed in giant molecular clouds when exposed to the radiation from newly formed OB-type stars in a central OB association can develop cometary head-tail morphology. Thus compression of non-collapsing clumps by shock waves driven by the warm surface gas could possibly drive the inner cores to instability and gravitational collapse, triggering star formation. The enhanced rate of star formation seen in CGs based on the detection of IRAS point sources having their spectral energy distributions characteristic of young stellar or protostellar objects towards the compact heads of CGs \citep{1993MNRAS.262..812B} could be a result of such triggered star formation.

Magnetic fields are thought to play a crucial role in the formation and subsequent evolution of molecular clouds \citep[e.g.,][]{1976ApJ...210..326M, 2000ApJ...540L.103B, 2012ARA&A..50...29C}. The effects of magnetic fields of various strengths and orientations on the formation and evolution of dense pillars and CGs at the boundaries of HII regions were investigated using 3D hydrodynamical simulations including photoionising radiative transfer simulations by \citet{2009MNRAS.398..157H} and \citet{2011MNRAS.412.2079M}. They found that a strong initial magnetic field is required to significantly alter the non-magnetized dynamics because the energy input from the photo-ionization is so strong that it controls almost the entire dynamics. In the cases of weak and medium field strengths, an initially perpendicular field is dragged and made to align with the pillar during the dynamical evolution of the CGs. A strong perpendicular field, however, remains in its original configuration during the dynamical evolution of the globules. 

Background starlight while passing through molecular clouds gets polarized due to the aligned, aspheric dust grains present in them. The polarisation is produced because of the selective extinction suffered by the light as it passes through the aspheric dust grains that are aligned to the magnetic field of the clouds. Though the exact alignment mechanism is still unclear \citep{2003JQSRT..79..881L, 2004ASPC..309..467R}, the selective  extinction due to aligned, aspherical dust grains would make the polarisation vectors to trace the direction of the plane-of-the-sky magnetic field of the molecular clouds. The contribution to the observed polarization depends on the amount of dust grains with sizes comparable to the wavelength of background starlight being observed \citep{1995ApJ...448..748G, 1996ASPC...97..325G}. Observationally, large grains are shown to exist inside dense clouds inferred through an increase in the value of the total-to-selective extinction ratio, R$_{V}\sim5$ \citep{1980ApJ...235..905W, 2003AJ....126.1888K, 2005ASPC..343..321W, 2010A&A...522A..84O}. Thus polarization observations in optical wavelengths are sensitive to the grains located at the periphery of the clouds where the extinction suffered by background starlight is typically low \cite[e.g.,][]{1995ApJ...448..748G, 2007ApJ...662.1014P, 2009MNRAS.398..394W, 2010ApJ...723..146F}. Therefore optical polarimetric method is useful to trace the magnetic field geometry of the outer layers of molecular clouds. 

Optical polarimetric observations of stars projected in the regions of some of the globules have been made earlier to understand the role played by the magnetic field in 1) the formation of head-tail morphology, 2) the orientation of outflows and binary components, if present, and 3) the star formation process in them.  Maps of magnetic field towards the \textit{IRAS} VELA Shell and relatively large star forming globules namely L810, B335 and ESO 210-6A have been made by \citet{2002ApJS..141..469P} and \citet{1987ApJ...319..842H} respectively. Magnetic field geometry of CG 22, CG 30-31 complex and CG 12 have been studied by \citet[][Paper I]{1996MNRAS.279.1191S}, \citet[][Paper II]{1999MNRAS.308...40B} and \citet[][Paper III]{2004MNRAS.348...83B} respectively.  In continuation to the above series of works, we mapped the magnetic field geometry of a relatively isolated cometary globule, LBN 437.

\section{Cometary Globule, LBN 437}

LBN 437 which is also known as Gal 96-15 \citep{1988BAAS...20..957O} is the edge of a molecular cloud known as Kh149 (Khavtassi, 1960) and is also located on the border of an HII region, S126 \citep{1959ApJS....4..257S}. The southern part of Kh149 is resolved into two condensation nuclei namely condensation A and B in $^{12}$CO and $^{13}$CO molecular line observations \citep{1994A&A...290..235O}. LBN 437 does not have a prominent tail. The nuclear region of LBN 437 is coincident with a reflection nebula \citep[DG187, ][]{1964AN....288...23D}, and contains a group of four H${\alpha}$ emission line sources namely, LkH$\alpha$ 233 (or V375 Lac), a Herbig A4e star \citep{2004AJ....127.1682H} and other fainter members namely LkH$\alpha$ 230, LkH$\alpha$ 231, and LkH$\alpha$ 232 \citep{2008Ap&SS.315..215M}. LkH$\alpha$ 233 with its surrounding nebulosity is one of the classic examples of bipolar appearance in Herbig Ae/Be stars \citep{1978MNRAS.182..687C}. LkH$\alpha$ 233 is the exciting source of Herbig-Haro (HH) objects \citep{1998A&A...336..535C} with collimated bipolar jets emanating at an angle of $68\degree$ with respect to the north. \citet{2006ApJ...653..657M} derived a mass of $\sim3M_{\odot}$, a luminosity of $L=\sim3~L_{\odot}$, and an age of $\sim1$ Myr for LkH$\alpha$ 233. These are the youngest stars found associated with Lacerta OB1 (Lac OB1) association \citep{2011AAS...21734028P}.

LBN 437 is considered to be at a distance of 460 pc on the basis of  spatial and kinematic coincidence with the Lac OB1 association \citep{1994A&A...290..235O}. But the distance to Lac OB1 association itself is highly uncertain. The estimated distances are in the range of $\sim360$ pc to $\sim600$ pc \citep{2009PASP..121.1045K}. Also, though LBN 437 is considered to be at 460 pc, the LkH$\alpha$ 233, is estimated to be at a distance of 880 pc. But the presence of reflection nebulosity surrounding LkH$\alpha$ 233 confirms their association. The distance of 880 pc was adopted from \citet{1978MNRAS.182..687C} who estimated the distance based on the inference that the B1.5 V star HD 213976, which is found projected on the cloud, has a distance modulus of 9.6 ($\sim$830 pc) and has negligible extinction ($A_{V}\sim0.42$) so should be foreground to the cloud. In majority of the subsequent works authors have adopted this distance to LkH$\alpha$ 233 \citep[e.g.,][]{1998A&A...336..535C, 2006ApJ...653..657M, 2007ApJ...670..499P}.  However, from the revised parallax measurements of this star ($3.08\pm0.56$ milliarcseconds) by \citet{2007A&A...474..653V}, the distance to this star is only  $\sim325^{+72}_{-50}$ pc \citep[also see][]{2009ApJ...697..824A}. This implies that the distance of LBN 437 is at least $\sim325$ pc if not more.

In this work we report the optical polarimetry of LBN 437 in order to study the magnetic field geometry at the periphery of the cloud. We measured optical polarisation of stars that are projected on LBN 437 and mapped the plane-of-the-sky magnetic field. This paper is organized in a manner that section \ref{sec:obs} describes the observations and the methods of data reduction. In section \ref{sec:res_dis}, we present our results and discussion. In the same section, we discuss the procedure used to determine distance to the cloud and to determine the foreground polarisation. Finally, we conclude our paper by summarising the results in section \ref{sec:conclude}.  


\section{OBSERVATIONS AND DATA REDUCTION}\label{sec:obs}

Polarimetric observations were carried using the Aries IMaging POLarimeter (AIMPOL) \citep{2004BASI...32..159R} mounted at cassegrain focus of the 104-cm Sampurnanand telescope of Aryabhatta Research Institute of Observational Sciences (ARIES), Nainital, India, coupled with TK 1024$\times$1024 $pixel^{2}$ CCD camera. AIMPOL consists of an achromatic half-wave plate (HWP) modulator and a Wollaston prism beam-splitter. The observations were carried out in standard Jhonson R filter having  $\lambda_{R_{eff}}$=0.630$\mu$m photometric band. Plate scale of CCD is 1.48 arcsec$/$pixel and field of view is $\sim 8$ arcmin in diameter. The full width at half maximum (FWHM) varies from 2 to 3 pixels. The Read out noise and gain of CCD are 7.0 $e^{-1}$  and 11.98 $e^{-1}$/ADU respectively. Table \ref{tab:obslog} gives the log of the observations.

\begin{table}
\caption{Log of observations in R filter ($\lambda_{R_{eff}}$=0.630$\mu$m).}\label{tab:obslog}
\begin{tabular}{p{1.3cm}p{6cm}}\hline
 Cloud ID          &  Date of observations (year, month,date)\\
\hline
 LBN 437           & 2011, November, 22, 23, 26; 2011, December, 20; 2012, October, 14, 19, 20\\
\hline
\end{tabular}
\end{table}

Fluxes of ordinary ({\it $I_{o}$}) and extraordinary ({\it $I_{e}$}) beams for all the observed sources with a good signal-to-noise ratio were extracted by standard aperture photometry using the IRAF package. The ratio {\it {R($\alpha$)}} is given by
 \begin{equation}
 R(\alpha) = \frac{\frac{{I_{e}}(\alpha)}{{I_{o}}(\alpha)}-1} {\frac{I_{e}(\alpha)} {I_{o}(\alpha)}+1} =  P cos(2\theta - 4\alpha)
\end{equation}
where {\it P} is the fraction of total linearly polarised light and $\theta$ is the polarisation angle of the plane of polarisation. Here {\it $\alpha$} is the position of the fast axis of HWP at $0\degree$, $22.5\degree$, $45\degree$ and $67.5\degree$ corresponding to four normalized Stokes parameters, respectively, q[R($0\degree$)], u[R($22.5\degree$)], $q_{1}$[R($45\degree$)] and $u_{1}$[R($67.5\degree$)]. We estimate the errors in normalised Stokes parameters ($\sigma_R$)($\alpha$)($\sigma_q$, $\sigma_u$, $\sigma_{q1}$, $\sigma_{u1}$) in per cent using the relation \citep{1998A&AS..128..369R}.
\begin{equation}
\sigma_R(\alpha)= \frac{\sqrt{N_{e}+N_{o}+2N_{b}}}{N_{e}+N_{o}}
\end{equation}
where $N_{o}$ and $N_{e}$ are the counts in ordinary and extraordinary beams, respectively, and $N_{b}$[= ({$N_{be}$}$+${$N_{bo}$})/2] is the average background counts around the extraordinary and ordinary rays. 

Zero polarization standard stars were observed during every run to check for any possible instrumental polarization. The typical instrumental polarisation is found to be less than $\sim0.1\%$. The instrumental polarisation of AIMPOL on the 104-cm Sampurnanand Telescope has been monitored since 2004 for various observing programs and found to be stable \citep[see ][]{2004BASI...32..159R, 2008MNRAS.388..105M, 2011MNRAS.411.1418E}. The reference direction of the polarizer was determined by observing polarized standard stars from \citet{1992AJ....104.1563S}. The results are presented in Table \ref{tab:std}.
We observed these un-polarised and polarised standards using the standard Johnson R filter having $\lambda_{R_{eff}}$=0.630$\mu$m. \citet{1992AJ....104.1563S} used Kron-Cousins R filter for the observations of the standard stars. Because the instumental polarization is found to be very low, we did not correct the observed degree of polarization. Neverthless, we found a good correlation of the observed values with the standard values (see Table 2). The zero point offset was corrected on every run using the offset seen between the standard position angle values given in \citet{1992AJ....104.1563S} and those obtained by us.

\begin{table}
\caption{Polarized standard stars observed in $R_c$ band.}\label{tab:std}
\begin{tabular}{lll}\hline
Date of     &P $\pm$ $\epsilon_P$ 	&  $\theta$ $\pm$ $\epsilon_{\theta}$  \\
Obs.		&(\%)            		& ($\degree$)\\    \hline
\multicolumn{3}{l}{{\bf HD 236633} ($^\dagger$Standard values: 5.38 $\pm$ 0.02\%, 93.04 $\pm$ 0.15$\degree$)}\\
22 Nov 2011 & 5.1 $\pm$ 0.1		& 93 $\pm$ 1 \\
26 Nov 2011 & 5.4 $\pm$ 0.1		& 92 $\pm$ 1 \\
20 Dec 2011	& 5.4 $\pm$ 0.2		& 93 $\pm$ 1 \\
14 Oct 2012	& 5.6 $\pm$ 0.2		& 99 $\pm$ 6 \\
19 Oct 2012	& 5.6 $\pm$ 0.2 	& 99 $\pm$ 5 \\
20 Oct 2012	& 5.4 $\pm$ 0.2 	& 99 $\pm$ 6 \\\hline
\multicolumn{3}{l}{{\bf HD 236954} ($^\ddagger$Standard values: 6.16 $\pm$ 0.17\%, 110.0 $\pm$ 0.8$\degree$)}\\
26 Nov 2011 & 5.9 $\pm$ 0.1   	& 111 $\pm$ 1\\
20 Dec 2011 & 6.2 $\pm$ 0.4   	& 111 $\pm$ 2\\        \hline
\multicolumn{3}{l}{{\bf BD$+$59$\degree$389} ($^\dagger$Standard values: 6.43 $\pm$ 0.02\%, 98.14 $\pm$ 0.10$\degree$)}\\
26 Nov 2011	& 7.0 $\pm$ 0.1   	& 98 $\pm$ 1 \\
20 Dec 2011 & 6.3 $\pm$ 0.1   	& 98 $\pm$ 1 \\
14 Oct 2012 & 6.5 $\pm$ 0.1 	& 105 $\pm$ 7\\        \hline
\multicolumn{3}{l}{{\bf HD 204827} ($^\dagger$Standard values: 4.89 $\pm$ 0.03\%, 59.10 $\pm$ 0.17$\degree$)}\\
20 Oct 2012 & 5.0 $\pm$ 0.2 	& 66 $\pm$ 7 \\        \hline
\end{tabular}

$\dagger$  Values in R band from \citet{1992AJ....104.1563S} \\
$\ddagger$ Values are in R band calculated using the Serkowski's law by taking polarization values of this star in V band given in \citet{2000AJ....119..923H} [Original reference \citep{1956ApJS....2..389H}] \\
\end{table}


\section{RESULTS AND DISCUSSION}\label{sec:res_dis}

\begin{figure}
\centering
\resizebox{8.1cm}{8.4cm}{\includegraphics{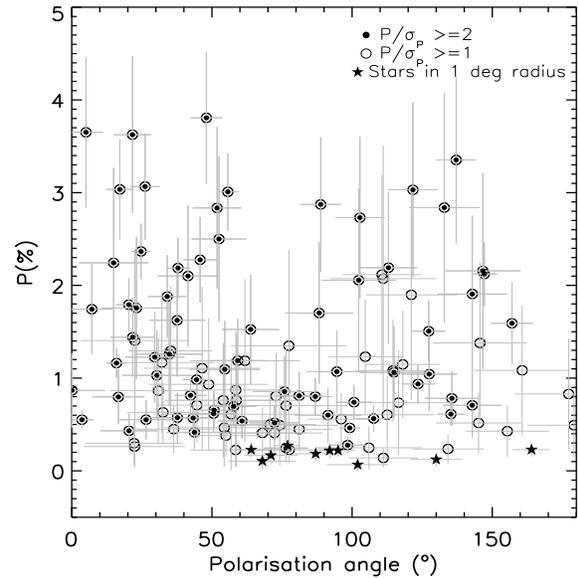}}
\caption {Degree of polarisation versus polarisation position angles of the stars projected towards the head region of LBN 437. Filled circles show the values with P/$\sigma_{p}$ $\geq$ 2 and open circles show the values with P/$\sigma_{p}$ $\geq$ 1. Polarisation values of stars obtained in 1$\degree$ radius around LBN437 are also shown using star symbols.}\label{fig:PvsPA_1_0.5}
\end{figure}

Results of our R-band polarimetry of 70 stars projected in the direction of LBN 437 are presented in Table \ref{tab:res70}. We have tabulated the results of only those sources for which the ratio of degree of polarisation (P\%) and error in the degree of polarisation ($\sigma_{p}$), P/$\sigma_{p}$ $\geq$ 2. Column 1 of Table \ref{tab:res70} shows the star identification in the increasing order of their Right Ascensions (RA). Columns 2 and 3 show the RA and the declination of the target objects. Columns  4 and 5 show the measured P (\%) values and the polarisation position angles ($\theta$ in degree). The position angles are measured from the north increasing towards the east. In Fig. \ref{fig:PvsPA_1_0.5}, we show the observed degree of polarisation versus the position angles of 116 stars (open circles) with P/$\sigma_{p}$ $\geq$ 1. Of these 70 stars with P/$\sigma_{p}$ $\geq$ 2 are identified using filled circles. Also shown in filled star symbols are the polarisation results observed by us for the ten stars that were obtained from a circular region of $1\degree$ radius around LBN 437 essentially to carry out foreground subtraction (see \S \ref{subsec:fg_sub}). The distances to these stars were estimated prior to making the observations using their parallaxes obtained from the catalogue produced by \citet{2007A&A...474..653V}.

\begin{table}
\begin{minipage}{80mm}
\caption{Polarisation results of 70 stars (with P/$\sigma_{p}$ $\geq$ 2) observed in the direction of LBN 437.}\label{tab:res70}
\begin{tabular}{llllr}  \hline
Star  & $\alpha$ (J2000)  & $\delta$ (J2000)  & P $\pm$ $\epsilon_P$ & $\theta$ $\pm$ $\epsilon_{\theta}$  \\ 
 Id  &($\degree$)&($\degree$)& (\%) &($\degree$) \\\hline  
1  &	338.304621  &	      $+$40.808308  &	0.5 $\pm$	0.2  &	72 $\pm$	9  \\
2  &	338.330770  &         $+$40.791203  &	0.6 $\pm$	0.2  &	50 $\pm$	8  \\
3  &	338.340637  &         $+$40.781494  &	1.1 $\pm$	0.4  &	95 $\pm$	11 \\
4  &	338.347652  &         $+$40.811455  &	3.1 $\pm$	0.6  &	26 $\pm$	5  \\
5  &	338.348541  &         $+$40.718292  &	3.1 $\pm$	0.4  &	56 $\pm$	4  \\
6  &	338.351013  &         $+$40.721653  &	0.4 $\pm$	0.1  &	43 $\pm$	9  \\
7  &	338.362701  &         $+$40.726276  &	0.7 $\pm$	0.2  &	58 $\pm$	8  \\
8  &	338.376831  &         $+$40.711460  &	0.7 $\pm$	0.2  &	16 $\pm$	9  \\
\multicolumn{5}{c}{}\\
9  &	338.382907  &         $+$40.682140  &	1.8 $\pm$	0.2  &	20 $\pm$	4  \\
10 &	338.385101  &         $+$40.758514  &	0.6 $\pm$	0.2  &	91 $\pm$	11 \\
11 &	338.385942  &         $+$40.725651  &	1.3 $\pm$	0.6  &	35 $\pm$	12 \\
12 &	338.408112  &         $+$40.696949  &	0.4 $\pm$	0.1  &	99 $\pm$	3  \\
13 &	338.408122  &         $+$40.696949  &	3.4 $\pm$	0.8  &	137$\pm$	7  \\
14 &	338.418640  &         $+$40.717636  &	0.8 $\pm$	0.4  &	0  $\pm$	11 \\
15 &	338.418665  &         $+$40.717644  &	1.1 $\pm$	0.5  &	55 $\pm$	12 \\
16 &	338.421631  &         $+$40.753284  &	1.1 $\pm$	0.3  &	127$\pm$	9  \\
\multicolumn{5}{c}{}\\
17 &	338.424561  &         $+$40.725250  &	2.1 $\pm$	0.8  &	112$\pm$	10 \\
18 &	338.425201  &         $+$40.705345  &	0.8 $\pm$	0.1  &	86 $\pm$	6  \\
19 &	338.428497  &         $+$40.708778  &	2.8 $\pm$	0.7  &	88 $\pm$	7  \\
20 &	338.431368  &         $+$40.818741  &	2.8 $\pm$	0.8  &	52 $\pm$	9  \\
21 &	338.439484  &         $+$40.689621  &	3.0 $\pm$	0.9  &	121$\pm$	8  \\
22 &	338.454315  &         $+$40.741238  &	1.5 $\pm$	0.4  &	157$\pm$	7  \\
23 &	338.455942  &         $+$40.674004  &	0.7 $\pm$	0.1  &	101$\pm$	6  \\
24 &	338.457550  &         $+$40.656029  &	0.8 $\pm$	0.3  &	136$\pm$	9  \\
\multicolumn{5}{c}{}\\
25 &	338.478546  &         $+$40.768497  &	2.7 $\pm$	0.8  &	102$\pm$	8  \\
26 &	338.481567  &         $+$40.731873  &	0.8 $\pm$	0.2  &	81 $\pm$	8  \\
27 &	338.482928  &         $+$40.659981  &	0.6 $\pm$	0.1  &	108$\pm$	6  \\
28 &	338.498152  &         $+$40.853371  &	0.6 $\pm$	0.1  &	4  $\pm$	4  \\
29 &	338.507139  &         $+$40.678768  &	1.1 $\pm$	0.4  &	115$\pm$	12 \\
30 &	338.524307  &         $+$40.649021  &	0.9 $\pm$	0.2  &	123$\pm$	6  \\
31 &	338.532696  &         $+$40.837280  &	2.1 $\pm$	0.9  &	102$\pm$	13 \\
32 &	338.558082  &         $+$40.829716  &	2.4 $\pm$	0.3  &	25 $\pm$	3  \\
\multicolumn{5}{c}{}\\
33 &	338.625755  &         $+$40.774902  &	0.7 $\pm$	0.2  &	50 $\pm$	7  \\
34 &	338.630520  &         $+$40.612507  &	1.5 $\pm$	0.3  &	127$\pm$	6  \\
35 &	338.632322  &         $+$40.749577  &	1.0 $\pm$	0.1  &	45 $\pm$	5  \\
36 &	338.643217  &         $+$40.572807  &	2.8 $\pm$	1.2  &	133$\pm$	13 \\
37 &	338.648243  &         $+$40.781261  &	3.8 $\pm$	0.7  &	48 $\pm$	6  \\
38 &	338.658751  &         $+$40.745464  &	2.0 $\pm$	0.8  &	143$\pm$	13 \\
39 &	338.659459  &         $+$40.808346  &	1.9 $\pm$	0.4  &	34 $\pm$	7  \\
40 &	338.666698  &         $+$40.598907  &	2.1 $\pm$	0.3  &	147$\pm$	4  \\
\multicolumn{5}{c}{}\\
41 &	338.667445  &         $+$40.626091  &	1.7 $\pm$	0.4  &	7  $\pm$	7  \\
42 &	338.688526  &         $+$40.602062  &	2.2 $\pm$	0.3  &	37 $\pm$	4  \\
43 &	338.689607  &         $+$40.807808  &	1.2 $\pm$	0.2  &	16 $\pm$	4  \\
44 &	338.689625  &         $+$40.714291  &	2.3 $\pm$	0.4  &	46 $\pm$	6  \\
45 &	338.697390  &         $+$40.748314  &	3.6 $\pm$	0.8  &	22 $\pm$	7  \\
46 &	338.702863  &         $+$40.768250  &	2.1 $\pm$	0.8  &	41 $\pm$	11 \\
47 &	338.704579  &         $+$40.590450  &	0.3 $\pm$	0.1  &	98 $\pm$	7  \\
48 &	338.710279  &         $+$40.732536  &	0.7 $\pm$	0.3  &	143$\pm$	10 \\
\multicolumn{5}{c}{}\\
49 &	338.711425 &         $+$40.751999  &	1.8 $\pm$	0.8  &	23 $\pm$	12  \\
50 &	338.750614 &         $+$40.749975  &	1.2 $\pm$	0.5  &	59 $\pm$	10  \\
51 &	338.750625 &         $+$40.749977  &	1.4 $\pm$	0.4  &	22 $\pm$	8   \\
52 &	338.754028 &         $+$40.740021  &	0.6 $\pm$	0.2  &	27 $\pm$	9   \\
53 &	338.754072 &         $+$40.754173  &	1.5 $\pm$	0.5  &	64 $\pm$	10  \\
54 &	338.755403 &         $+$40.735828  &	2.2 $\pm$	1.0  &	15 $\pm$	12  \\
55 &	338.755413 &         $+$40.735808  &	3.0 $\pm$	0.5  &	17 $\pm$	4   \\
56 &	338.758214 &         $+$40.716522  &	1.7 $\pm$	0.8  &	88 $\pm$	12  \\
\hline
\end{tabular}
\end{minipage}
\end{table}
\begin{table}
\centering
\begin{minipage}{80mm}
\contcaption{}
\begin{tabular}{llllr}  \hline
Star  & $\alpha$ (J2000)  & $\delta$ (J2000)  & P $\pm$ $\epsilon_P$ & $\theta$ $\pm$ $\epsilon_{\theta}$  \\\hline  
57  &	338.761076  &         $+$40.748074 &	0.3 $\pm$	0.1 &	76 $\pm$	8  \\
58  &	338.764024  &         $+$40.608459 &	0.9 $\pm$	0.3 &	76 $\pm$	11 \\
59  &	338.770245  &         $+$40.732365 &	2.5 $\pm$	0.8 &	52 $\pm$	9  \\
60  &	338.785987  &         $+$40.693169 &	1.2 $\pm$	0.4 &	29 $\pm$	10 \\
61  &	338.789490  &         $+$40.771175 &	3.7 $\pm$	0.8 &	5  $\pm$	6  \\
62  &	338.791325  &         $+$40.678619 &	1.6 $\pm$	0.7 &	38 $\pm$	10 \\
63  &	338.791888  &         $+$40.746410 &	0.6 $\pm$	0.1 &	135$\pm$	4  \\
64  &	338.793716  &         $+$40.682663 &	0.8 $\pm$	0.1 &	42 $\pm$	6  \\
\multicolumn{5}{c}{}\\
65	&	338.799687  &         $+$40.738823 &	0.6 $\pm$	0.2 &	38 $\pm$	11 \\
66	&	338.842246  &         $+$40.701893 &	0.4 $\pm$	0.1 &	20 $\pm$	10 \\
67	&	338.847091  &         $+$40.770603 &	0.5 $\pm$	0.1 &	61 $\pm$	9  \\
68	&	338.861710  &         $+$40.758518 &	1.0 $\pm$	0.4 &	30 $\pm$	12 \\
69	&	338.871566  &         $+$40.733170 &	0.6 $\pm$	0.1 &	43 $\pm$	5  \\
70	&	338.882723  &         $+$40.711555 &	2.2 $\pm$	1.0 &	147$\pm$	13 \\
\hline	   	   
\end{tabular}	   	   
\end{minipage}
\end{table}

\subsection{Distance to the cloud and foreground interstellar polarisation}\label{sec:fg_dist}

The observed polarisation of any star background to a cloud has two components. One is due to the dust located in the cloud and the other is due to the dust in the interstellar medium (ISM) located between the observer and the cloud. Therefore, in order to obtain the true polarisation (magnetic field) geometry of a cloud, it is essential to subtract the component due to the ISM vectorially from the observed polarisation values. The contribution to the polarisation values due to the foreground ISM can be estimated by measuring the polarisation values of the foreground stars for which the distance estimations are already available. Subtraction of these values from the observed results can give the true estimate of polarisation due to the cloud material alone. But subtraction of the foreground contribution is possible only when the distance to the cloud is known to avoid the use of background stars in the subtraction.

\subsubsection{Distance to LBN 437}\label{subsec:dist}

\begin{figure}
\resizebox{8cm}{8.5cm}{\includegraphics{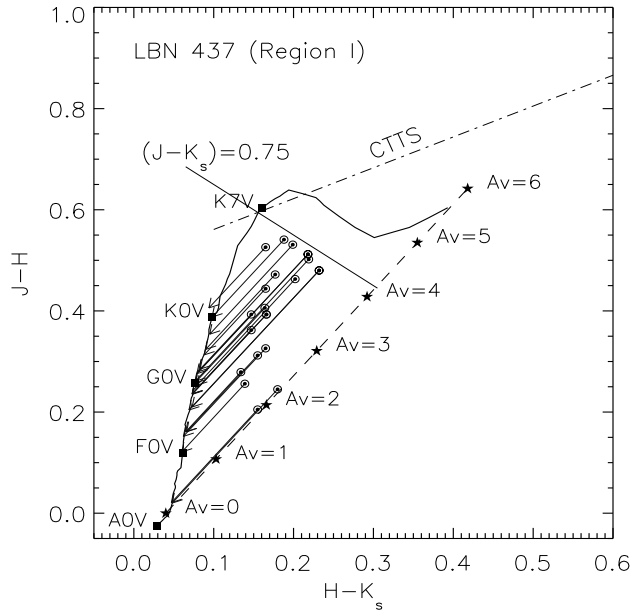}}
\caption{The ($J-H$) vs. ($H-K_{s}$) CC diagram drawn for stars (with $A_{V}\geq1$) from Region I which contains LBN 437 to illustrate the method. The solid curve represents locations of  unreddened main sequence stars. The reddening vector for an A0V type star drawn  parallel to the \citet{1985ApJ...288..618R} interstellar reddening vector is shown by the dashed line. The locations of the main sequence stars of different spectral types are marked  with square symbols. The region to the right of the reddening vector is known as the NIR  excess region and corresponds to the location of PMS sources.  The dash-dot-dash line represents the loci of unreddened CTTSs \citep{1997AJ....114..288M}. The open circles represent the observed colours and the arrows are drawn from the observed to the final colours obtained by the method for each star.}\label{fig:CC}
\end{figure}

We used the near-IR photometric method presented by \citet{2010A&A...509A..44M}, which utilizes the vast homogeneous $JHK_{s}$ photometric data produced by the Two Micron All Sky Survey \citep[2MASS, ][]{2003yCat.2246....0C} available for the entire sky, to determine distance of LBN 437. Here we present a brief discussion of the method\footnote{For a more rigorous discussion on the errors and limitations of the method, refer \citet{2010A&A...509A..44M}}. The method uses a technique by which  spectral classification of stars lying towards the fields containing the clouds can be made into main sequence and giants. The observed ($J-H$) and ($H-K_{s}$) colours of the stars with ($J-K_{s}$)$\leq0.75$ in ($J-H$) vs. ($H-K_{s}$) colour-colour (CC) diagram are de-reddened simultaneously using trial values of $A_{V}$  and a normal interstellar extinction law \citep{1985ApJ...288..618R}. The best fit of the de-reddened colours to the intrinsic  colours giving a minimum value of $\chi^{2}$ then yields the corresponding spectral type and $A_{V}$ for the star. The main sequence stars, thus classified, are plotted in an $A_{V}$ versus distance diagram to bracket the cloud distance. The entire procedure is depicted in Fig. \ref{fig:CC} where we plot the near infra-red CC diagram for the stars (with $A_{V}\geq1$) chosen from the region I towards the direction of LBN 437. The arrows are drawn from the  observed data points (open circles) to the corresponding de-reddened colours estimated using the method. The values of maximum extinction that can be measured using the method are those for A0V type stars ($\approx4$ magnitude). The extinction traced by the stars will decrease as we move towards more late type ones. 

Sub-dividing the field containing a cloud is always better in order to avoid confusions that could arise due to any erroneous classifications of giants as dwarfs. While the rise in the extinction due to the presence of a cloud should occur almost at the same distance in all the fields, if the whole cloud is  located at the same distance, the wrongly classified stars in the sub-fields would show high extinction not at same but at random  distances. In the case of clouds with smaller angular sizes, other clouds  that are located spatially closer and show similar radial velocities could be selected. Here the assumption is that the clouds that are spatially closer and share similar velocities are located almost at similar distances.

\begin{figure}
\centering
\resizebox{8cm}{8cm}{\includegraphics{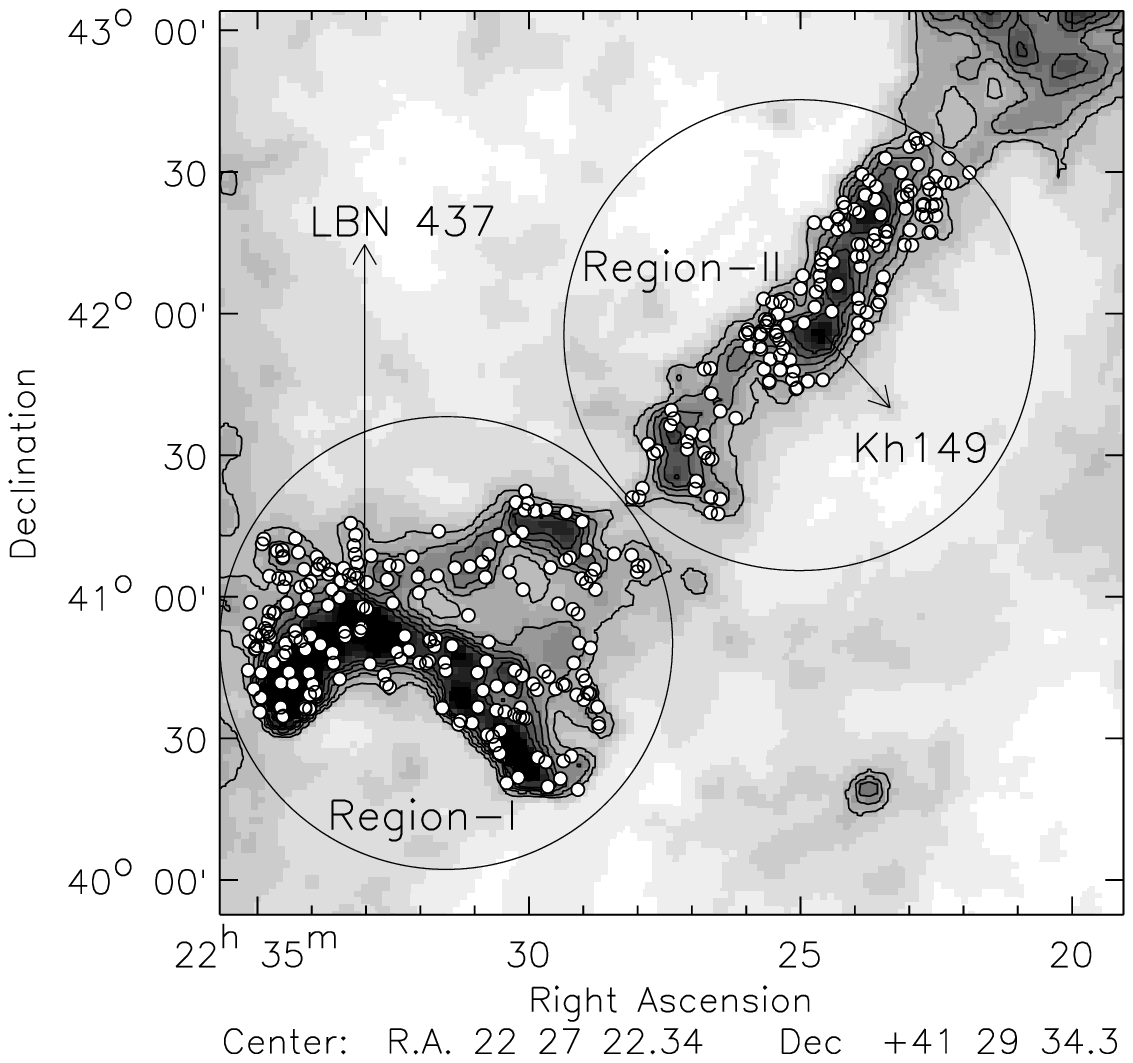}}
\resizebox{8cm}{8cm}{\includegraphics{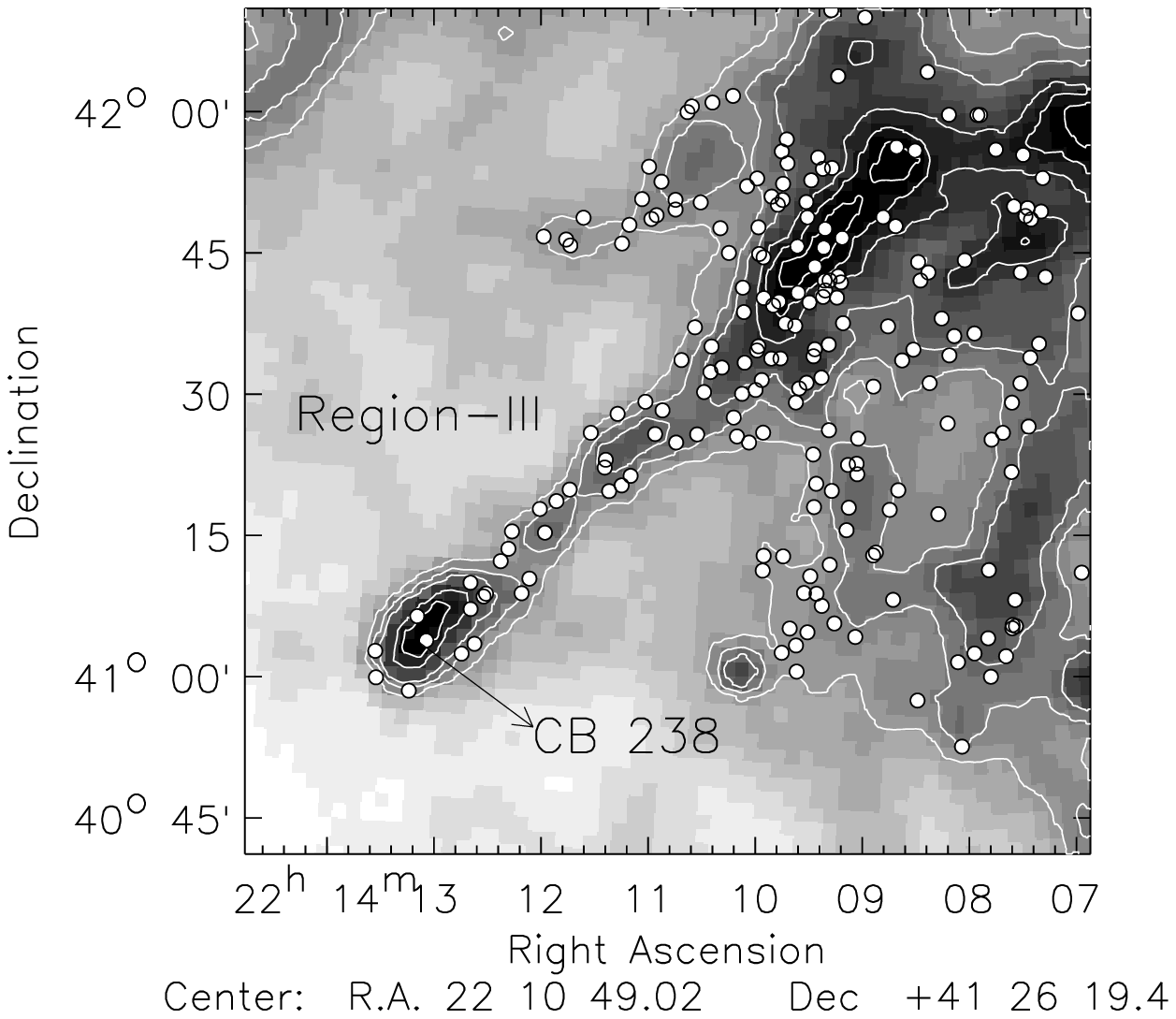}}
\caption{{\bf Upper panel:} The $3\degree\times3\degree$ IRAS 100$\mu$m image of the field containing LBN 437. The Regions I and II include LBN 437 and Kh 149 respectively. {\bf Lower panel:} The $1.5\degree\times1.5\degree$ IRAS 100$\mu$m image of the field containing CB 238. The stars selected for determining distance to the cloud are identified using circles on both the panels.}.\label{fig:dist_images}
\end{figure}
\begin{figure}
\centering
\resizebox{8cm}{13cm}{\includegraphics{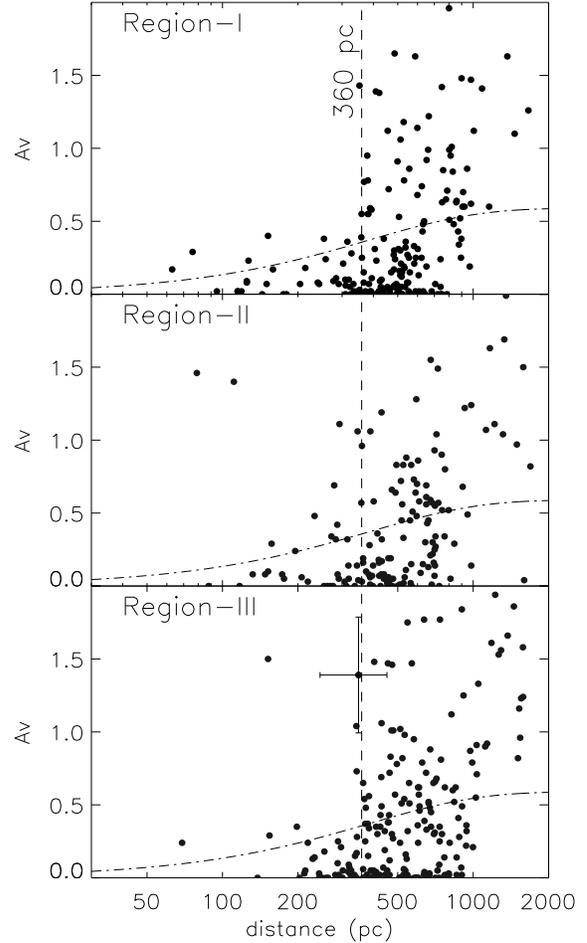}}
\caption{The $A_{V}$ vs. $d$ plots for the stars from the Regions I, II, and III towards LBN 437 shown using filled circles. The dashed vertical line is drawn at 360 pc inferred from the procedure described in \citet{2010A&A...509A..44M}. The dash-dotted curve represents  the increase in the extinction towards the Galactic latitude of $b=-15\degree$ as a function of distance produced from the expressions given by \citet{1980ApJS...44...73B}. A typical error-bar is shown on a point at a distance of $\sim$360 pc}.\label{fig:individualdist}
\end{figure}

In the upper panel of Fig. \ref{fig:dist_images} we show the two regions that are selected towards LBN 437 to determine the distance. The regions I and II include LBN 437 and Kh 149 respectively. The additional field, region III (shown in the lower panel of Fig. \ref{fig:dist_images}) includes CB 238, a relatively small and isolated cloud located at $\sim4\degree$ west of LBN 437 \citep[$l=93.47\degree, b=-12.63\degree$; ][]{1988ApJS...68..257C}. We included this cloud because the radial velocity of CB 238 \citep[0.2 km$~s^{-1}$, ][]{1988ApJS...68..257C} and LBN 437 \citep[$-$0.2 km$~s^{-1}$, ][]{1994A&A...290..235O} are found to be similar. Furthermore,  by comparing the images shown in the upper and lower panel of the Fig.  \ref{fig:dist_images} it is quite apparent that the finger like structure of CB 238 is also pointing almost exactly to the same direction as is in the case of LBN 437. The same exciting source might be responsible for the current structure of both the clouds. 

In Fig. \ref{fig:individualdist}, we present the $A_{V}$ vs. $d$ plot for the stars from the regions I, II and III. The dash-dotted curve represents  the increase in the extinction towards the Galactic latitude of $b=-15\degree$ as a function of distance produced from the expressions given by \citet{1980ApJS...44...73B}\footnote{The expression for the obscuration as a function of distance from the Sun ($R$) and the galactic latitude ($b$) that \citet{1980ApJS...44...73B} used in their model for the Galaxy was $A(R) = A_{\infty}(b)~[1-exp(-\sin b/H)R]$. The $A_{\infty}(b)=A_{\infty}(90^{\degree})~\csc b$, where $A_{\infty}(90^{\degree})=0.15$. They assumed an exponential variation of the density of the obscuring material with height above the galactic plane, $\rho=\rho_{o}~exp(-z/H)$, where the scale height $H$ is taken as 100 pc.}. When compared to the values of extinction expected towards the latitude of $-15\degree$, a sudden increase in the extinction is noticeable in all the three panels at a distance close to 360 pc. At this distance, LBN 437 would be at a height of 93 pc above the galactic plane, close to the scale height of 100 pc assumed for the distribution of the obscuring material by \citet{1980ApJS...44...73B} in their Galaxy model. In general, for line of sight close to the plane of the Milky Way and for distances upto a few kilo parsec from the Sun, the visual extinction along the path length is taken as $\sim1.8$ mag/kpc \citep{2003dge..conf.....W}. But because the scale height of the exponential distribution of obscuring material was taken as 100 pc  by \citet{1980ApJS...44...73B}, the dash-dotted line, showing the extinction, tends to become more flatter with the distance towards the direction of LBN 437.

In order to determine the distance at which the sudden jump in the extinction is occurring, we first grouped the stars into distance bins of $bin~width = 0.18\times distance$. The centres of each bin are separated by half of the bin width. Because of very few stars at smaller distances, the mean value of the distances and the $A_{V}$ of the stars in each bin were calculated by taking 1000 pc as the initial point and proceeded towards smaller distances. The mean distance of the stars in the bin at which a  significant drop in the mean of the extinction occurred was taken as the distance to the cloud and the average of the uncertainty in the distances of the stars in that bin was taken as the final uncertainty in distance determined by us for the cloud. The error bars on the mean $A_{V}$ values are calculated using standard deviation of $A_{V}$ values corresponding to each bin. The vertical dashed line in $A_{V}$ vs. $d$ plots, used to mark the cloud distance,  is drawn at distance deduced from the above procedure. We determined a distance of $360\pm65$ pc to both LBN 437 and CB 238.
 
\begin{figure}
\resizebox{8.5cm}{7cm}{\includegraphics{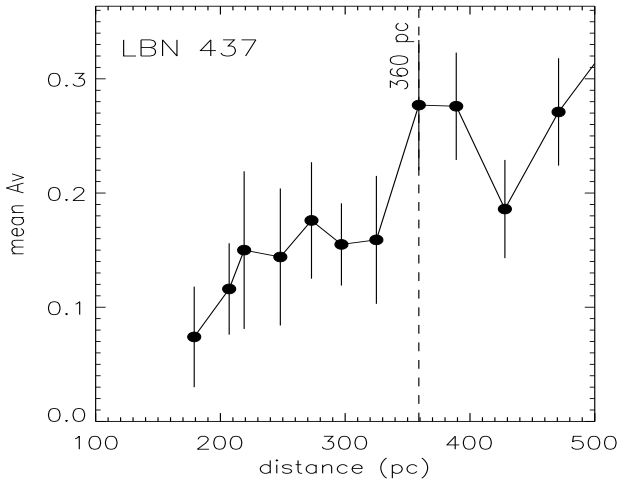}}
\caption{The mean values of $A_{V}$ vs. the mean values of distance plot for LBN 437 produced using the the procedure discussed in \citet[][see the text for a brief description]{2010A&A...509A..44M}.  The distance at which the first sharp increase in the mean value of extinction occur is taken as the distance to the cloud. The error bars on the mean $A_{V}$ values are calculated from the standard deviation of $A_{V}$ values in each bin.\label{fig:hist}}
\end{figure}

Our distance estimate of 360 pc to LBN 437 is consistent with the distance recently estimated by \citet{2009ApJ...697..824A} based on the photometric distances determined for HD 213976, LkH$\alpha$ 231 and  LkH$\alpha$ 232. For HD 213976, they derived a distance of $383^{+153}_{-109}$ pc by comparing the absolute $J-$band brightness of this star with the mean absolute  $J-$band brightness of Upper Scorpious objects. For LkH$\alpha$ 231 and  LkH$\alpha$ 232, they showed that the observed $J-$ band magnitudes of these objects are consistent with the similar spectral type (K2-K5) stars in the Taurus when placed at 325 pc. Our distance of 360 pc to LBN 437 is similar to the distance of Lac OB1 ($368 \pm 17$ pc) estimated by \citet{1999AJ....117..354D}. At a distance of 360 pc, LBN 437 is located $\sim90$ pc away from the Galactic plane.


\subsubsection{Subtraction of interstellar polarisation}\label{subsec:fg_sub}

\begin{figure}
\centering
\resizebox{9.3cm}{9cm}{\includegraphics{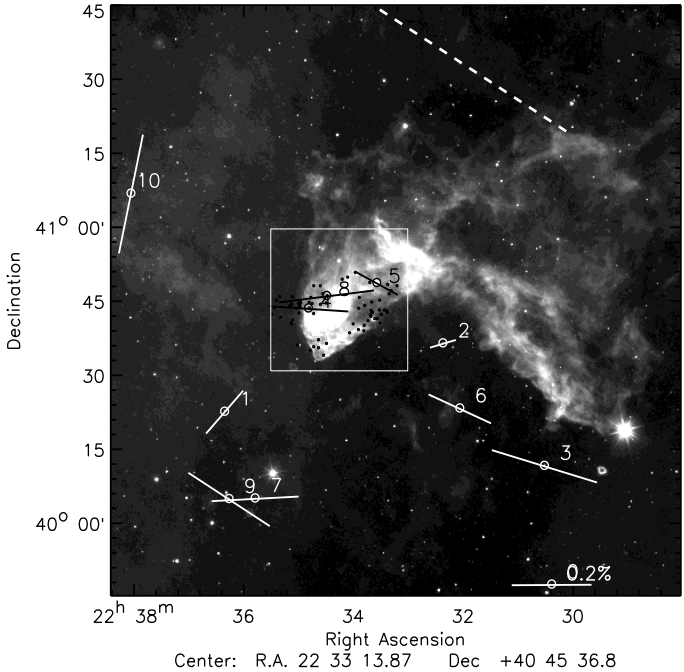}}
\caption{Positions of ten stars that are obtained within 1$\degree$ radius around LBN 437 with available parallax values in \citet{2007A&A...474..653V}. The distances to these ten stars range from $190$ pc to $385$ pc. The polarisation vectors (white) corresponding to these stars are over-plotted on the $2\degree\times2\degree$ Wide-Field Infrared Survey Explorer (WISE) 12$\mu$m image of the field containing LBN 437. The filled circles in black (enclosed in square box) identify the target sources that we observed towards the head of LBN 437. The broken line in white color is drawn parallel to the Galactic plane at $b=-15\degree$.}\label{fig:DSSvectors}
\end{figure}

In order to subtract the polarisation component due to the foreground material from our observed values, we made a search for stars that are located within a circular region of $1\degree$ radius about LBN 437 and have their parallax values measured by the Hipparcos satellite. We obtained ten stars for which the parallax measurements are available in the catalogue produced by \citet{2007A&A...474..653V}. We rejected those that are classified as emission line stars, stars in a binary or multiple system or are peculiar according to the information provided by the Simbad.  From the catalogue, we selected only those stars for which the values of the ratio of the error in parallax and the parallax measurements are $\leq$ 0.5. The ten stars thus selected are listed in Table \ref{tab:fg} ordered according to increasing distance. We then carried out polarimetric observations of these stars in R-band using the AIMPOL. The polarisation vectors corresponding to these stars are over-plotted on the $2\degree\times2\degree$ Wide-Field Infrared Survey Explorer (WISE) 12$\mu$m image of the field containing LBN 437 as shown in Fig. \ref{fig:DSSvectors}. The filled circles in black identify the target sources that we observed towards the head (enclosed in square box) of LBN 437. The broken line in white is drawn parallel to the Galactic plane at $b=-15\degree$.

\begin{figure}
\resizebox{8.7cm}{12.5cm}{\includegraphics{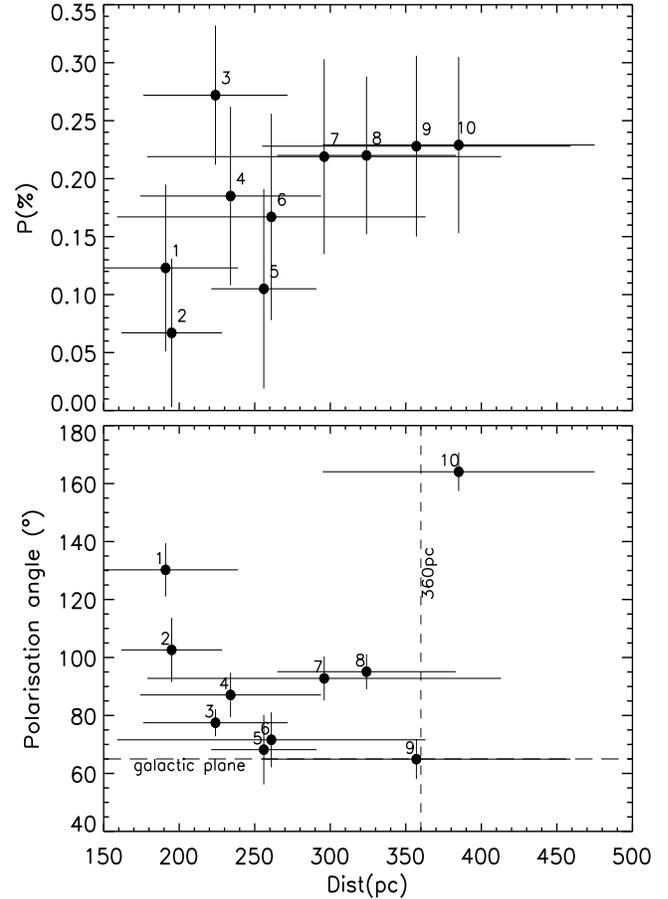}}
\caption{{\bf Upper panel:} Degree of polarisation versus distance for the nine foreground stars and one star (star \#10) at 385pc obtained in a circular region of $1\degree$ radius about LBN 437. {\bf Lower panel:} Polarisation position angle versus their distances for the same stars. The position angle of the Galactic plane at $b=-15\degree$ and the distance of LBN 437 are marked using broken lines. The stars are identified using the same numbers as given in Table \ref{tab:fg}.}\label{fig:PvsDist1degree}
\end{figure}

The polarimetric results of these foreground stars are shown in Fig. \ref{fig:PvsDist1degree}. The distances to these ten stars range from $190$ pc to $385$ pc. There are nine stars that are at distances less than 360 pc. Of these nine, we excluded the star \#2 as the degree of polarization is both very low and below the instrumental polarization. For the purpose of subtracting the foreground polarisation, we estimated the weighted mean of polarisation values of remaining eight foreground stars and calculated the Stokes parameters. The weighted mean values of the degree of polarization and the position angles are found to be 0.2\% and 85$\degree$.  Using these values, we calculated the mean Stokes parameters $Q_{fg} (=P\cos 2\theta$) and $U_{fg} (=P\sin 2\theta$) as -0.176 and 0.028 respectively. Then we calculated the Stokes parameters, $Q_{\star}$ and $U_{\star}$, of the target sources. The Stokes parameters $Q_{c}$ and $U_{c}$ representing the foreground corrected polarisation of the target stars are calculated using
\begin{equation} \label{qu_star_ism}
Q_{c}=Q_{\star} - Q_{fg},\\
U_{c}=U_{\star} - U_{fg}
\end{equation} 
Corresponding foreground corrected degree of polarisation $P_{c}$ and position angle $\theta_{c}$ of the target stars are calculated using the equations
\begin{equation} \label{ppa_star_ism}
P_{c}=\sqrt{(Q_{c})^2+(U_{c})^2},\\
\theta_{c}=0.5\times tan^{-1}\left(\frac{U_{c}}{Q_{c}}\right)
\end{equation}
Histogram of the polarisation position angles of the stars with $P/\sigma_{p} \geq$ 1, after subtracting the foreground contribution is shown in Fig. \ref{fig:CorrectP_PA_sig_1_0.5}. Also shown using open circles are the degree of polarisation versus position angles of the target stars. Filled circles show the polarisation values with $P/\sigma_{p}\geq 2$. No significant changes are noticed in the observed polarisation results before and after the subtraction of the foreground contribution because the polarization due to the foreground dust component is very small, at the level of $\sim0.2\%$. This implies that the polarisation values obtained for the majority of the target stars projected on the head of LBN 437 are mainly caused due to the dust component that is associated with the cloud. This is also evident from the fact that the degree of polarization of HD 214243 which is located just in front of LBN 437 shows a very low polarization. The observed low polarization of this star could be either due to the star being at relatively high galactic latitude or due to the local field direction being pointed along the line of sight. 

\begin{table}
\caption{Our R-band polarisation results for 9 foreground stars.}\label{tab:fg}
\begin{tabular}{llllll}\hline
Id&Star	Name	&V   	& P $\pm$ $\epsilon_P$ & $\theta$ $\pm$ $\epsilon_{\theta}$  &D$^{\dagger}$\\ 
  &				&(mag)	& (\%) 			&($\degree$)	&(pc) \\\hline
1 &HD 214283	&9.0 	& 0.12$\pm$0.07 & 130$\pm$ 9&	192\\   
2 &HD 213659	&8.0 	& 0.07$\pm$0.06 & 103$\pm$11&	196\\   
3 &HD 213421	&8.2 	& 0.27$\pm$0.06 &  78$\pm$ 5&	224\\   
4 &HD 214022	&8.5 	& 0.19$\pm$0.08 &  87$\pm$ 8&	234\\   
5 &HD 213835	&6.5 	& 0.11$\pm$0.09 &  68$\pm$12&	256\\  
6 &BD$+$39$\degree$4868	&9.5 	& 0.17$\pm$0.09 &  72$\pm$10&	262\\   
7 &BD$+$39$\degree$4890	&9.5 	& 0.22$\pm$0.08 &  93$\pm$ 8&	297\\   
8 &HD 213976	&7.0 	& 0.22$\pm$0.07 &  95$\pm$ 6&	325\\   
9 &HD 214243	&8.3 	& 0.23$\pm$0.08 &  65$\pm$ 7&	357\\
10&HD 214524    &7.5    & 0.23$\pm$0.07& 164$\pm$ 7&   385\\
\hline
\end{tabular}

$^{\dagger}$ distances are estimated using the Hipparcos parallax measurements taken from \citet{2007A&A...474..653V}.
\end{table}


\subsection{Magnetic field geometry of LBN 437} 

\begin{figure*}
\resizebox{13cm}{13cm}{\includegraphics{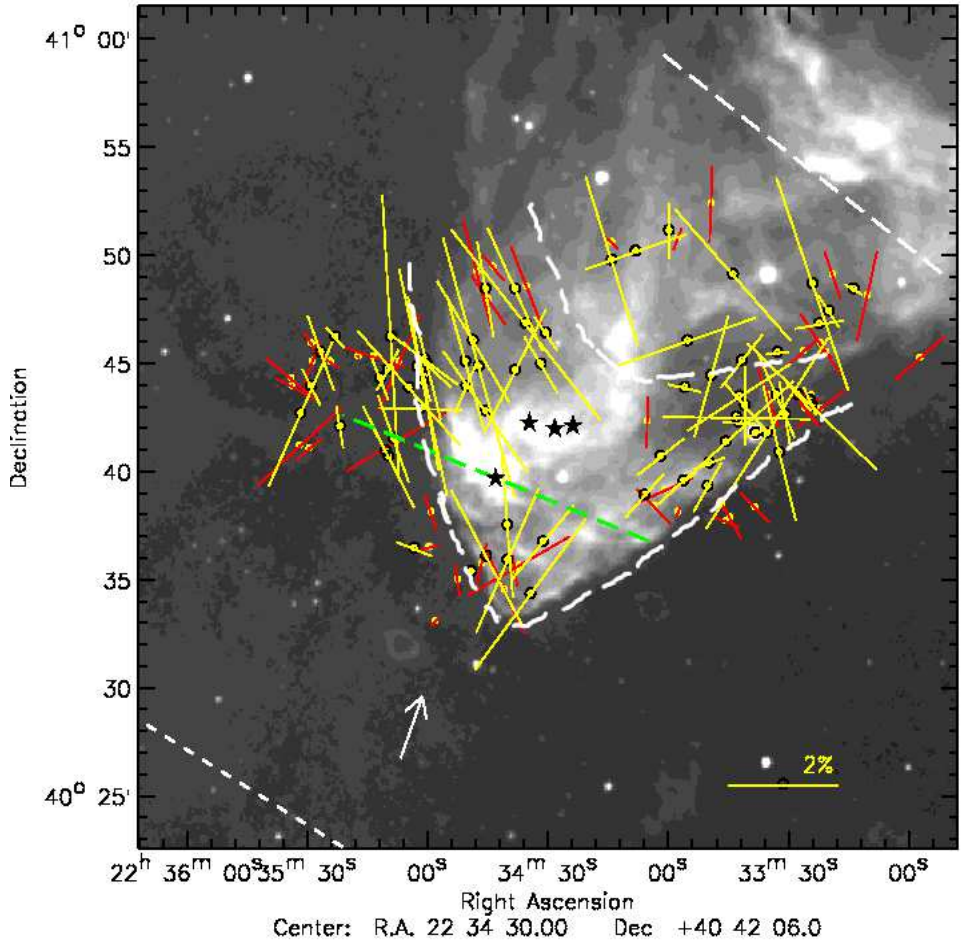}}
\caption{Polarisation vectors plotted on $0.65\degree$ $\times$ $0.65\degree$ WISE 12$\mu$m image of LBN 437 after subtracting the foreground polarisation contribution. Red vectors correspond to the values with P/$\sigma_{p}$ $\geq$ 1 and the yellow vectors show the values with P/$\sigma_{p}$ $\geq$ 2. The dashed line vector in green color shows the direction of outflow from the star LKH$\alpha$ 233. The white dashed line shows the Galactic plane at $b=-15\degree$. Positions of LKH$\alpha$ 230, LKH$\alpha$ 231 and LKH$\alpha$ 232 are identified using star symbols in black colour. The thick broken curve represent the inferred magnetic field orientation from the polarisation vectors. An arrow is drawn to show the spatial direction of 10 Lac. A vector with 2$\%$ polarisation is shown as reference.}\label{fig:PolVectWISE12}
\end{figure*}

\begin{figure}
\centering
\resizebox{8.5cm}{9cm}{\includegraphics{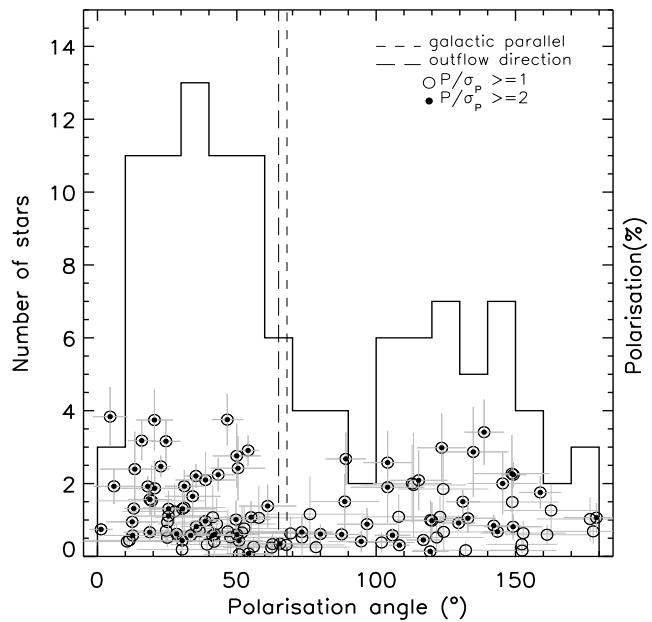}}
\caption{Histogram of the polarisation position angles of the stars with P/$\sigma_{p}$ $\geq$ 1 after subtracting the foreground polarisation contribution. Also shown is the distribution of degree of polarisation versus position angles of target stars. Filled circles show the values with P/$\sigma_{p}$ $\geq$ 2 and open circles show the values with P/$\sigma_{p}$ $\geq$ 1. Broken lines show the position angles of galactic plane and the direction of outflow from the star LKH$\alpha$ 233.}\label{fig:CorrectP_PA_sig_1_0.5}
\end{figure}

The resultant foreground corrected values of $P_{c}$ and $\theta_{c}$ are over-plotted on $0.65\degree$ $\times$ $0.65\degree$ WISE 12$\mu$m image as shown in Fig. \ref{fig:PolVectWISE12}. The length of the vectors corresponds to the degree of polarisation and their orientation corresponds to the position angle measured from the north and increasing eastward. The red vectors correspond to the values with $P/\sigma_{p}\geq1$ and the yellow vectors show the values with $P/\sigma_{p}\geq2$. The mean values of the degree of polarisation and the position angles after foreground subtraction are found to be 1.6$\%$ and $61\degree$ respectively. From Fig. \ref{fig:CorrectP_PA_sig_1_0.5}, if we select the sources with their degree of polarisation $\geq$ 1\%, evidently, there exist two groups. The first group has polarisation position angles distributed from $\sim0\degree$ to $\sim65\degree$ and the second group has position angles distributed from $\sim100\degree$ to $\sim160\degree$. It can be noticed in Fig. \ref{fig:PolVectWISE12} that the two components are not distributed uniformly over the cloud. While the first group is found to be concentrated towards the eastern parts of the cloud head, the second group is dominant towards the western parts of the globule. The mean and the standard deviation of the first and the second groups are 37$\degree$ and 17$\degree$, and 125$\degree$ and 17$\degree$ respectively. Evidently, the magnetic field lines traced by our polarisation measurements are following the curved structure of the head. 

In order to study the effects of the presence of magnetic field on the dynamics of the dense neutral gas, \citet{2011MNRAS.412.2079M} included the magnetic fields of various strengths and orientations to the 3D hydrodynamic simulations which included photoionising radiative transfer also. Prior to this work, \citet{2009MNRAS.398..157H} studied the photoionisation of a dense clump of gas in 3D with an initially uniform magnetic field. They found that the presence of a strong magnetic field can significantly alter the evolution of a photoionised globule. Even without magnetic fields it was shown that the evolution of a photoionising globule proceeds in two processes. The first, radiation-driven implosion \citep[RDI; ][]{1989ApJ...346..735B} and the second, the acceleration due to the rocket effect \citep{1955ApJ...121....6O} producing elongated structures. The RDI provides initial compression to the neutral gas until it comes in equilibrium with the ionised gas while the rocket effect accelerates the globule away from the ionising source producing elongated structures like elephant trunks that are seen towards the periphery of a number of HII regions.

\citet{2011MNRAS.412.2079M} considered three different magnetic field strengths in their work. The weak, medium and strong fields corresponding to 18, 53 and 160$\mu$G respectively, that are oriented perpendicular to the direction of propagation of the ionising radiation. Whether a field is strong or weak is determined by its dynamical importance set by the plasma parameter $\beta~(=8\pi p_{g}/B^{2}$) which is the ratio between the thermal to the magnetic pressure. From their study they found that the initially perpendicular weak field orientation significantly got altered due to the RDI process. The weak field is swept into alignment with the pillar structure by the dynamics of RDI and rocket effect. In the case of perpendicular field orientation with medium strength, the field structure got altered only slightly whereas in the case of perpendicular field orientation with strong strength, hardly any change was noticed.

The luminous stars associated with the Lac OB1 association are considered to be responsible for the cometary shape of LBN 437 \citep{1994A&A...290..235O}. Of these the brighest star, namely 10 Lac which is of O9V spectral type, is located to the south-west of LBN 437. The line joining the positions of 10 Lac and LBN 437 subtends an angle of $\sim150\degree$ with respect to the north. This is taken as the direction of the ionising photons. To compare our results with those from the simulations, it is essential to know the initial magnetic field orientation (with respect to the direction of the ionising radiation) that was prevailing in the globule before it was subjected to the external ionising radiation. \citet{2009ApJ...704..891L} showed that there exists a significant alignment between the mean magnetic field direction in the cloud cores traced by the Hertz and the SCUpol and those of their surrounding inter cloud media traced by the optical polarimetery. They argued that due to the flux freezing, the cores could acquire the magnetic field orientation of the inter cloud medium. Using the star, HD 214243 which is located close to and in front of the globule, we infer that the ambient magnetic field orientation at the location close to the cloud is $\sim65\degree$ which is also parallel to the Galactic plane at $b=-15\degree$. If the same field was inherited by LBN 437, the initial magnetic field orientation prior to the action of the ionising source was $\sim65\degree$ which is almost perpendicular to the direction of the ionising radiation from 10 Lac.


The cometary head of LBN 437 might have been created when the ionising radiation from the Lac OB1 members interacted with the cloud. Due to the subsequent dynamical evolution of the globule, the initially perpendicular magnetic field lines got dragged away from the ionising source to follow the curvature of the globule head (as illustrated in Fig. \ref{fig:PolVectWISE12} using the white broken curves). However, the effect of the ionising radiation might not have been adequate enough to align the field lines completely along the direction of the ionising radiation as is seen towards a number of regions e.g. M16 \citep{2007PASJ...59..507S}, CG22 \citep{1996MNRAS.279.1191S} etc. and as shown in the simulations \citep{2009MNRAS.398..157H, 2010MNRAS.403..714M, 2011MNRAS.412.2079M}. Therefore, either the strength of the initial magnetic field was sufficiently strong enough to resist a complete alignment of the field lines with the ionising radiation or that the ionising source(s) was relatively far away that the extent of its effect on the globule was not very severe. Adopting a distance of 368 pc to both Lac OB1 association \citep{1999AJ....117..354D} and to 10 Lac (assuming it to be a member of Lac OB1), the spatial distance between 10 Lac and LBN 437 is estimated to be $\sim12$ pc which is much larger than those considered in the simulation studies in which the ionising source is kept at distances of less than 1 pc from the globule initially \citep{2009MNRAS.398..157H, 2010MNRAS.403..714M, 2011MNRAS.412.2079M}. The Herbig Ae star, LkH$\alpha$ 233, might have formed due to the gravitational collapse triggered by the compression due to the RDI process. The age of this star is estimated to be $\sim$1 Myr \citep{2006ApJ...653..657M} which is sufficiently larger than the typical time scale for the ionisation front to pass through the cloud \citep{1989ApJ...346..735B, 2011MNRAS.412.2079M}. According to the simulation results of \citet{2009MNRAS.398..157H}, the typical time scale for the RDI process to occur in a globule of $\sim15$ $M_{\odot}$ at a distance of less than one pc from the ionising source is about 50,000 yr. 

Using near-IR imaging polarimetery aided by the adaptive optics, \citet {2004Sci...303.1345P} showed the presence of a narrow, unpolarized dark lane consistent with an optically thick circumstellar disk blocking the direct light from LkH$\alpha$ 233. This dark lane is found to be almost perpendicular to the axis of the outflow which is oriented at an angle of $68\degree$ with  respect to the north \citep{2008A&A...483..199M}. The results obtained by \citet {2004Sci...303.1345P} was explained using models with an inclination angle of $80\degree$ implying that the outflow from LkH$\alpha$ 233 is almost on the plane-of-the-sky. \citet{2006ApJ...637L.105M}, using the polarised thermal dust emission from MHD simulations of protostellar collapse and outflow, showed that the alignment of an outflow with the magnetic field depends on the strength of the magnetic field inside the cloud core (at 1000 AU scale). They found that a magnetic field strength of 80$\mu$G could make the outflow align preferably with the mean polarisation vector of the cloud core. We find that the initial ambient magnetic field ($65\degree$) and the outflow directions are almost parallel to each other. This suggests that at the time of the triggered formation of LkH$\alpha$ 233, and the subsequent outflow phase of the protostar, the magnetic field might not have got modified inside the cloud allowing for an alignment between the outflow and  ambient magnetic field directions. Also, the magnetic field strength inside the cloud might be greater than  80$\mu$G which comes in the range of medium and strong regimes considered by  \citet{2011MNRAS.412.2079M} in their simulations. A very little modification of the original field orientation towards LBN 437 also supports for a relatively strong magnetic field that could be prevailing there. 

LBN 437 was observed in molecular lines of CO, NH$_{3}$ and H$_{2}$CO \citep{1994A&A...290..235O}. In H$_{2}$CO(1$_{11}$-1$_{10}$), $^{13}$CO(1-0) and (2-1) line observations they found that the head part is having an elongated configuration of size $\sim1.6\times0.4$ pc$^{2}$ (corrected for the 360 pc distance). The NH$_{3}$(1,1) and (2,2) inversion line observations of the globule head showed the existence of a cold and dense core of elliptical shape. The major axis of the elliptical core is found to be oriented perpendicular to both the ambient magnetic field and the outflow from LkH$\alpha$ 233. LBN 437 do not possess a well defined tail. The head-tail morphology of LBN 437 resembles more like a comma-structure rather than a cometary shape as is often found in the case of CGs (e.g., CG 1, CG 12 etc.). In LBN 437, the material behind the head is also found to be oriented parallel to the ambient magnetic field direction. It would be interesting to carry out polarimetric observations of the background stars in near-IR and optical wavelengths to infer the magnetic field geometry at the inner high density regions and the tail part of the globule. 

A large scale magnetic field mapping of the region containing cometary globules is required to understand the processes involved in the evolution of such globules. Such results could be used to compare with those from the simulations. Also required are the near-IR and submillimeter polarimetry of these globules to get the inner field orientations which could allow us to understand the relationship between the outflow and the magnetic field orientations better.

\section{CONCLUSIONS}\label{sec:conclude}

In this work we present the results of optical linear polarisation measurements of seventy stars projected towards a cometary globule, LBN 437. The main results obtained in this study are given below.
\begin{enumerate}
\item We determined a distance of 360$\pm$65 pc to two cloud, namely LBN 437 and CB 238, using near-IR photometric method. 

\item To interpret the magnetic field geometry of the globule in relation with the globule structure and outflow from the Herbig Ae star, LkH$\alpha$ 233, we subtracted the contribution of the foreground dust component from our results using polarimetric observations of ten stars (obtained from a circular region of radius $1\degree$ about LBN 437) with distances already known from the parallax measurements of the Hipparcos. The star, HD 214243, located at a distance of $357\pm102$ pc shows a polarisation position of $65\degree$. We considered the position angle of this star to be the orientation of the ambient and the cloud magnetic field prior to the cloud being subjected to the ionising radiation. Interestingly, this field is found to be oriented parallel to the Galactic plane at the latitude of the cloud.

\item LBN 437, thus, presents a scenario where the direction of the initial field lines prior to it being affected by the ionising radiation was perpendicular to the direction of the ionising radiation. We found that the magnetic field lines in the globule are curved in a manner that they follow the curvature of the globule head. The possible explanation is that the magnetic field lines might have got dragged due to the radiation from the same ionising source that was responsible for the cometary shape of the cloud. 

\item The outflow direction from LkH$\alpha$ 233 is found to be parallel to both the ambient magnetic field and the Galactic plane at the location of the cloud.

\end{enumerate}
The magnetic field geometry of the inner high density region towards the head, the tail part and the region covering Kh 149 would be very useful to understand the formation history of LBN 437 and its surrounding environment.

\section{ACKNOWLEDGEMENT}
This research has made use of the SIMBAD database, operated at CDS, Strasbourg, France. We also acknowledge the use of NASA's \textit{SkyView} facility (http://skyview.gsfc.nasa.gov) located at NASA Goddard Space Flight Center. C.W.L was supported by Basic Science Research Program though the National Research Foundation of Korea (NRF) funded by the Ministry of Education, Science, and Technology (2010-0011605).

\bibliographystyle{mn2e}
\bibliography{Gal96ref}

\label{lastpage}
\end{document}